\documentstyle[epsfig]{mn}

\title[Radiation Driven Implosion]{Radiation Driven Implosion of
Molecular Cloud Cores}

\author[O.~Kessel-Deynet, A.~Burkert]{O.~Kessel-Deynet, A.~Burkert\\
        Max-Planck-Institut f\"ur Astronomie, K\"onigstuhl 17, D-69117
        Heidelberg, Germany}

\begin{document}

\maketitle
\begin{abstract}
We present the first three-dimensional calculations of Radiation Driven Implosion of Molecular Clouds including self-gravity and ionization. We discuss the effects of initial density perturbations on the dynamics of ionizing globules and show that the onset of gravitational collapse can be significantly delayed for a multiple of the implosion timescale. We demonstrate that Radiation Driven Implosion could be an efficient process for injecting disordered kinetic energy into molecular clouds in the vicinity of massive stars.   
\end{abstract}

\section{Introduction}

Massive stars have a strong impact on the evolution of their parental
molecular clouds (MCs). Their fast winds and
their ionizing radiation cause an important feedback of energy and
momentum into the circumstellar environment.

One of the most interesting effects massive stars could have on MCs is
``sequential star formation'' \cite{elmegreen77}. When massive stars heat
their environment to equilibrium temperatures $\simeq$ 10\,000~K via
their ionizing UV radiation, the ionized region is 
divided from the surrounding molecular gas by a very thin transition layer
called 'ionization front'. Due to the high pressure in the hot,
ionized gas, a shock front is driven into the cooler molecular gas
which sweeps up a thin, dense layer of shocked material between it and
the ionization front. This layer is likely to be gravitationally
unstable and thus may form a new generation of stars on timescales of several million years. 

Molecular clouds are highly irregular,
with observed structures from large scales down to even the smallest
observable scales. There is ongoing debate whether the morphology
of molecular clouds can be described by a fractal. Broad linewidths in
molecular clouds also suggest irregular supersonic motion, which
indicates the presence of turbulence. In the presence of ionizing radiation,
theoretical  models predict that density enhancements will likely be excavated
by the ionization front \cite{elmegreen95}. They will subsequently protrude
into the ionized gas, and are directly exposed to the ionizing radiation.  

The dynamical behaviour (Radiation Driven Implosion, RDI) in such
exposed globules was studied numerically by Sandford, Whitaker \& Klein~
\shortcite{sandford82,sandford84}, who pointed out that the mass
distribution of stars formed in a perturbed molecular cloud affected
by RDI may sensitively depend on the spatial scales of the perturbations.
A similar approach was followed by Lefloch \&
 Lazareff~\shortcite{lefloch94}. In their models, the time evolution
can be subdivided into two stages: the implosion phase, lasting for
about 100~kyrs, and, if gravitationally stable, a subsequent
quiet, quasi-static cometary phase, in which the globule is slowly
eroded by ionization on timescales of several Myrs. Through
semi-analytical models, Bertoldi~\shortcite{bertoldi89} and Bertoldi \&
McKee~\shortcite{bertoldi90} constrained the ranges for mass and ionizing
flux in which gravitational instability in the compressed globules
should be expected. More recently, Hollenbach \& Bally (1998) and
L\'{o}pez-Martin et al. (2001) investigated the photoevaporation of
dense clumps of gas by an external source. The hydrodynamics of photoionized
columns in the Eagle Nebula has been modeled by  Williams et al. (2001).

Sugitani, Tamura \& Ogura~\shortcite{sugitani99,sugitani00} observed a
sample of 89 bright-rimmed clouds (BRCs), apparently objects with properties as
mentioned above, and found an excess in the luminosities and
luminosity-to-cloud ratios of embedded IRAS sources when compared to
sources in isolated globules, which is an indication for enhanced star
formation in their BRCs. Near-IR imaging of 44 BRCs
revealed that young stellar objects seem to be aligned
along the axis towards the ionizing cluster. There is even evidence for an
age gradient with older stellar objects closer to the OB cluster and
the youngest objects well inside the globules, aligned with the IRAS sources.
This result is consistent with sequential star formation while the shock
front advances further and further into the molecular cloud. 

The interplay between dynamics and self-gravity in exposed globules has not yet been treated in detail. Using the new, fast SPHI
method described in Kessel \& Burkert~\shortcite{kessel00}, we
present 3D calculations of imploding clumps including self-gravity.
We find evidence that initial small-scale density perturbations can lead to a
substantial stabilization against gravitational collapse during the
implosion phase and the subsequent cometary phase, even if the 
parameters of the globules suggest them to be highly unstable
according to the analytical estimations of Bertoldi \& McKee~\shortcite{bertoldi90}.  

\section{Initial conditions}

\subsection{Density distribution} \label{chapt:inconds}

The initial conditions are static solutions to the equation of
Emden~\shortcite{emden07}, called ``Bonnor-Ebert spheres''
\cite{bonnor56}. In these spheres, which are supported by an external
pressure at the truncation radius $r_{\rm t}$, the gravitational force is everywhere balanced by the
pressure gradient, which is vanishing at the center. The scale-free
form of the differential equation describing this situation is 
\begin{equation} \label{eq:bonnor}
  \frac{1}{\xi^2} \frac{{\rm d}}{{\rm d} \xi} \left( \xi^2 \frac{{\rm d}\psi}{{\rm
  d}\xi} \right) = \exp(-\psi).
\end{equation}
The relation between the parameters $\xi$ and $\psi$ on the one hand
and physical quantities density $\rho$ and radius $r$ on the other hand is given by
\[
  \rho = \lambda \exp(- \psi), \;\;\; r = \beta^{1/2} \lambda^{-1/2} \xi,
\]
where $\lambda$ is the maximum density for $\psi = 0$ and
\[
  \beta = \frac{kT}{4 \pi G \mu}
\]
is the temperature of the isothermal gas. Here, $k$ is Boltzmann's
Constant, $G$ the Gravitational Constant and $\mu$ the mean molecular weight.
                
 Stability analysis shows that there is a critical
value $\psi_{\rm c}$ for the truncation radius of the sphere. Below $\psi_{\rm c}$, the sphere
is stable against radial density perturbations. Above, it commences gravitational collapse on a few gravitational timescales if 
perturbed. At $\psi_{\rm c}$, the sphere contains about one Jeans mass of
matter, and the density contrast from the center to the pressure boundary is
$\sim 14$. Since we don't expect such supercritical cores to survive long
enough in molecular clouds to be influenced by ionizing radiation in relevant
number, we instead restrict ourselves to the critical case with the
truncation radius $\psi_{\rm t}=\psi_{\rm c}$. This enables us to study
whether compression by ionizing radiation can force one initial Jeans
mass into gravitational collapse. If not, we can conclude that induced
collapse for cases with masses less than a Jeans mass should also be
highly improbable. Additional simulations of initially Jeans-stable
clumps with $\psi_{\rm t} < \psi_{\rm c}$ will be studied in a
subsequent paper.

The maximum particle density of H atoms in the cores is assumed to be $\lambda=1000\;{\rm cm}^{-3}$ and the equilibrium temperature in the neutral cloud material $T_{\rm c}=10\;{\rm K}$ to mimic the typical conditions observed in molecular clouds.

The integration volume has the shape of a cube, which contains the Bonnor-Ebert
sphere, embedded in a surrounding gas of constant density $n_{\rm s}$. This
gas provides the boundary pressure for the sphere. The diameter of the sphere is $0.8$ times the size of the box. Note that this configuration alone would be
unstable against gravitational collapse due to the additional mass of
the surrounding gas, in which
pressure gradient and gravitational force are not in balance. However, the
timescale of this collapse is much longer than the dynamical timescale which
is imposed on the system by the large sound velocity of about $13$~km\,s$^{-1}$
in the hot, ionized medium. The setup chosen is not thought to provide a
stable starting configuration. Our intention is rather to start with a
reasonable density distribution, in which the initial core contains about one
Jeans mass.  

We allow material to stream freely away from the initial integration volume,
i.e. no boundary conditions are imposed.

\subsection{Initial density perturbations}

To investigate the effects of density irregularities, we impose in some calculations
Gaussian random perturbations on the initial density
distribution as described by Klessen \& Burkert~\shortcite{klesbur}. The density distribution is determined by the power
spectrum $P(k) \propto k^\alpha$. This is achieved by applying the
Zel'Dovich~\shortcite{zeldovich70} approximation. Besides the
unperturbed case without and with self-gravity (in the following cases
A and B), we ran perturbed cases with $\alpha=-3$ (more power on large
scales, case C) and $\alpha=-1$ (more power on small scales, case
D). The amplitude of the perturbations was chosen to be highly
non-linear, with a mean relative variance of $\sigma_{\rm r}=\sigma/\rho=0.5$.

\subsection{Parameters of the radiation field}

The ionizing flux falls perpendicular onto one of the cube
surfaces with a rate of photons per time and area, $F_0$. We specify the
''strength'' $C$ of the radiation field as the ratio between the cube size
and the path length $l$ which the radiation can penetrate into the surrounding
gas via an r-type ionization front. This length is given by the balance between
$F_0$ and the recombinations along $l$:
\begin{equation}
  F_0 = \int_0^l n(s)^2 \alpha_{\rm B} ds = n_{\rm s}^2 \alpha_{\rm B} l,
\end{equation}
where $\alpha_{\rm B}$ denotes the recombination coefficient adopting the ''on
the spot'' approximation~\cite{baker62}. In other words, the parameter
$C$ determines how much of the density enhancement is excavated by the ionization front before it turns from weak r-type into weak d-type.
From that moment on the medium dynamically reacts to the
pressure difference across the ionization front. For the calculations in this
paper, we chose $C=.97$.

\section{Results}

\subsection{Time evolution without self-gravity: case A} \label{sect:caseA}

\begin{figure*}
  \caption{Model A (see 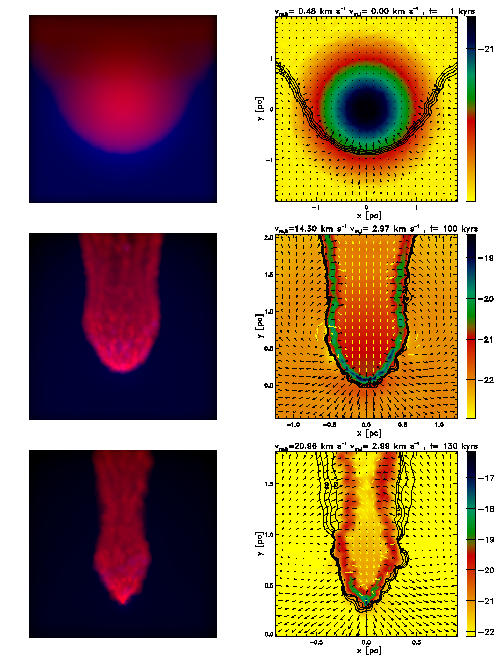). Radiation impinging from the bottom. Left
  column: 3D projection. Ionized gas is
  shown in blue, neutral gas in red. Brightness is a measure for the
  column density. Right column: cuts through the model along the
  symmetry axis. Color coded: log
  density in g/cm$^3$. Contour lines: ionization fraction. Arrows: velocity
  field. $v_{\rm m,h}$ ($v_{m,l}$) corresponds to longest black
  (bright) arrow. Evolution time
  indicated in the upper right corner.} \label{fig:A1} 
\end{figure*}
\begin{figure*}
  \caption{Model A (continued, see 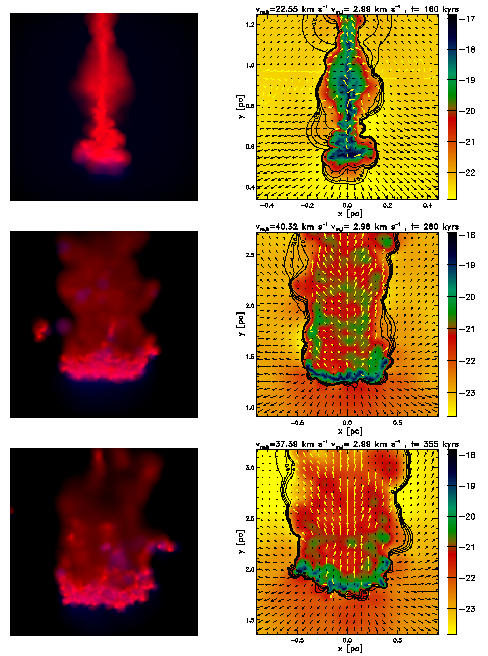). Radiation impinging from the bottom. Left column: 3D projection. Ionized gas is
  shown in blue, neutral gas in red. Brightness is a measure for the
  column density. Right column: cuts through the model along the
  symmetry axis. Color coded: log
  density in g/cm$^3$. Contour lines: ionization fraction. Arrows: velocity
  field. $v_{\rm m,h}$ ($v_{m,l}$) corresponds to longest black
  (bright) arrow. Evolution time
  indicated in the upper right corner.} \label{fig:A2} 
\end{figure*}
\begin{figure*}
  \caption{Model C (see 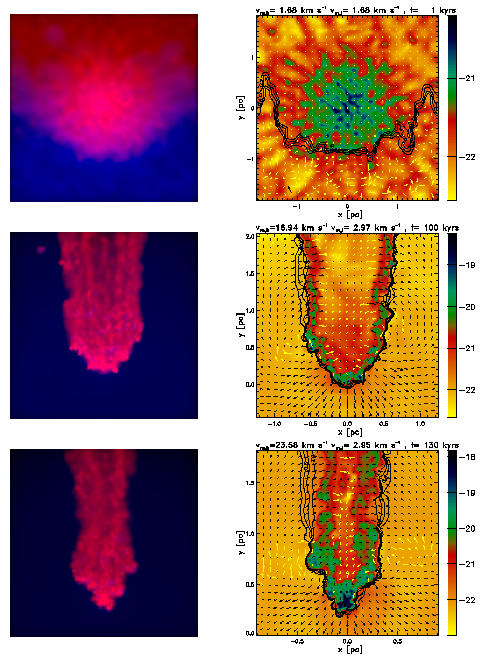). Radiation impinging from the bottom. Left column: 3D projection. Ionized gas is
  shown in blue, neutral gas in red. Brightness is a measure for the
  column density. Right column: cuts through the model along the
  symmetry axis. Color coded: log
  density in g/cm$^3$. Contour lines: ionization fraction. Arrows: velocity
  field. $v_{\rm m,h}$ ($v_{m,l}$) corresponds to longest black
  (bright) arrow. Evolution time
  indicated in the upper right corner.} \label{fig:C1} 
\end{figure*}
\begin{figure*}
  \caption{Model C (continued, see 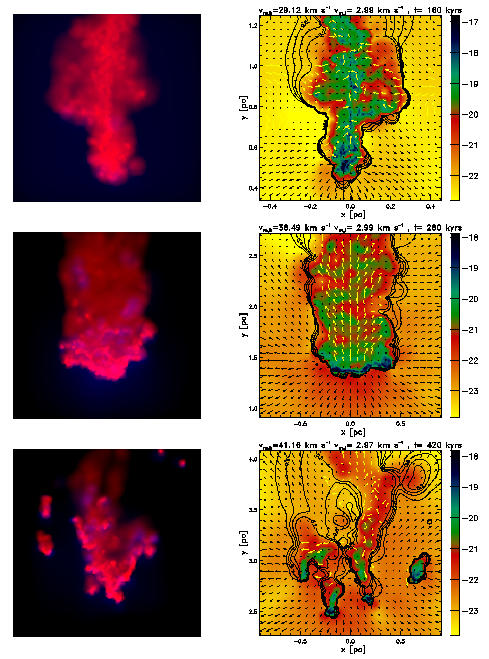). Radiation impinging from the bottom. Left column: 3D projection. Ionized gas is
  shown in blue, neutral gas in red. Brightness is a measure for the
  column density. Right column: cuts through the model along the
  symmetry axis. Color coded: log
  density in g/cm$^3$. Contour lines: ionization fraction. Arrows: velocity
  field. $v_{\rm m,h}$ ($v_{m,l}$) corresponds to longest black
  (bright) arrow. Evolution time
  indicated in the upper right corner.} \label{fig:C2} 
\end{figure*}

The time evolution in case A (Figs.~\ref{fig:A1},\ref{fig:A2}) is analogous to the models of
Lefloch \& Lazareff~\shortcite{lefloch94}. One can distinguish between four
different phases:
\begin{enumerate}
\item Immediately after the ionizing flux is switched on, a weak
R-type ionization front is driven supersonically into the
medium. After $\sim 1.5$~kyrs, the denser part of the cloud is
excavated and surrounded by a region of ionized gas. The ionization
front stalls and becomes nearly sonic. It converts into a weak D-type
front. The gas can now react dynamically to the pressure imbalance between ionized and unionized gas (Fig.~\ref{fig:A1}, first panels).
\item In the second phase, this imbalance drives a shock front into the cloud with velocities $\simeq 5 {\rm km/s}$. The shock accumulates a thin layer of dense shocked material (Fig.~\ref{fig:A1}, second panels). It thereby converges towards the axis of the finger-like structure (``globule'') due to the curvature of the ionization front.
\item After 130 Myrs, the globule reaches a stage of maximum
compression (Fig.~\ref{fig:A1}, third panels). Caused by the overshoot in pressure, the cloud starts reexpanding and 
\item enters a fourth phase, in which gas is
being ionized continuously away from the surface facing the ionizing
source (Fig.~\ref{fig:A2}).
This leads to a mass loss of $\simeq 10^{-4} \; {\rm M}_\odot/{\rm yr}$. Gas is leaving the ionization
front with velocities of $\simeq 13 \; {\rm km/s}$, causing a rocket
effect which accelerates the gas in the tip into the direction of
photon flow. During this acceleration, gas is collected inside the globule in a lid-shaped
feature facing the ionizing source. It is subject to dynamical
instabilities which lead to the disintegration of the cloud after
$\sim 600$ kyrs. The remaining fragments are then ionized completely
after a few 10 kyrs.
  
\end{enumerate}

\subsection{Time evolution with self-gravity, no initial density perturbations: case B}

\begin{figure*}
    \epsfig{file=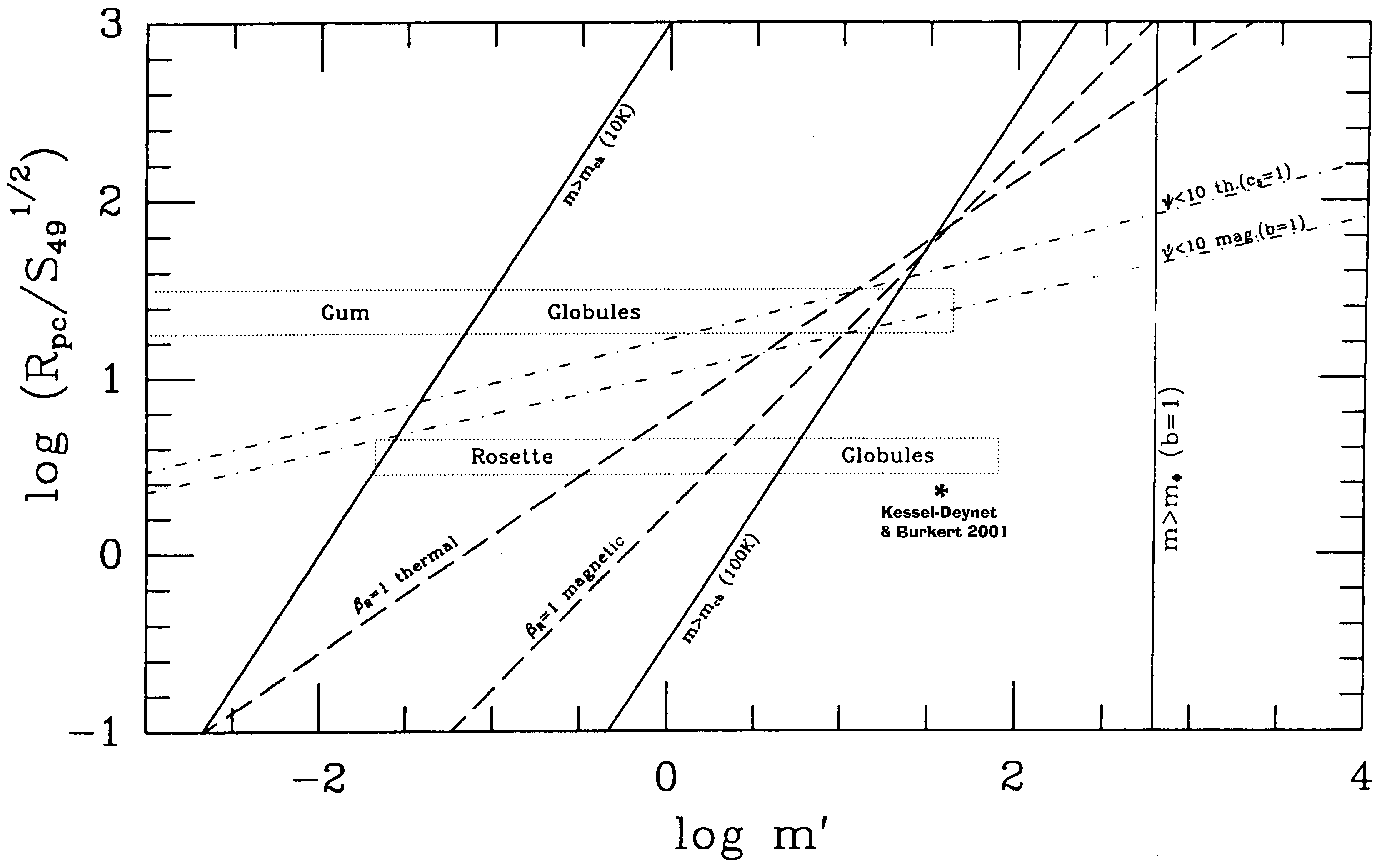,width=17cm}
  \caption{  Figure 14 reprinted from Bertoldi \&
McKee~\shortcite{bertoldi90}, which shows the parameter space of
  stratified cometary globules. $R_{\rm pc}$ is the distance from the
  ionizing source in parsec, $S_{49}$ the number of ionizing photons
  emitted in units of $10^{49} / {\rm s}$, and $m'$ the mass of the
  globule in units of $M_\odot$. The solid lines are thresholds of
  gravitational stability for different temperatures. Globules
  positioned to the right of these lines should be gravitationally
  unstable. The models presented here are indicated by an
  asteric. They lie far to the right of the $T=10\;{\rm K}$ stability
  line. They should thus collapse after stratification.
  } \label{fig:bertoldi} 
\end{figure*}

\begin{figure}
    \epsfig{file=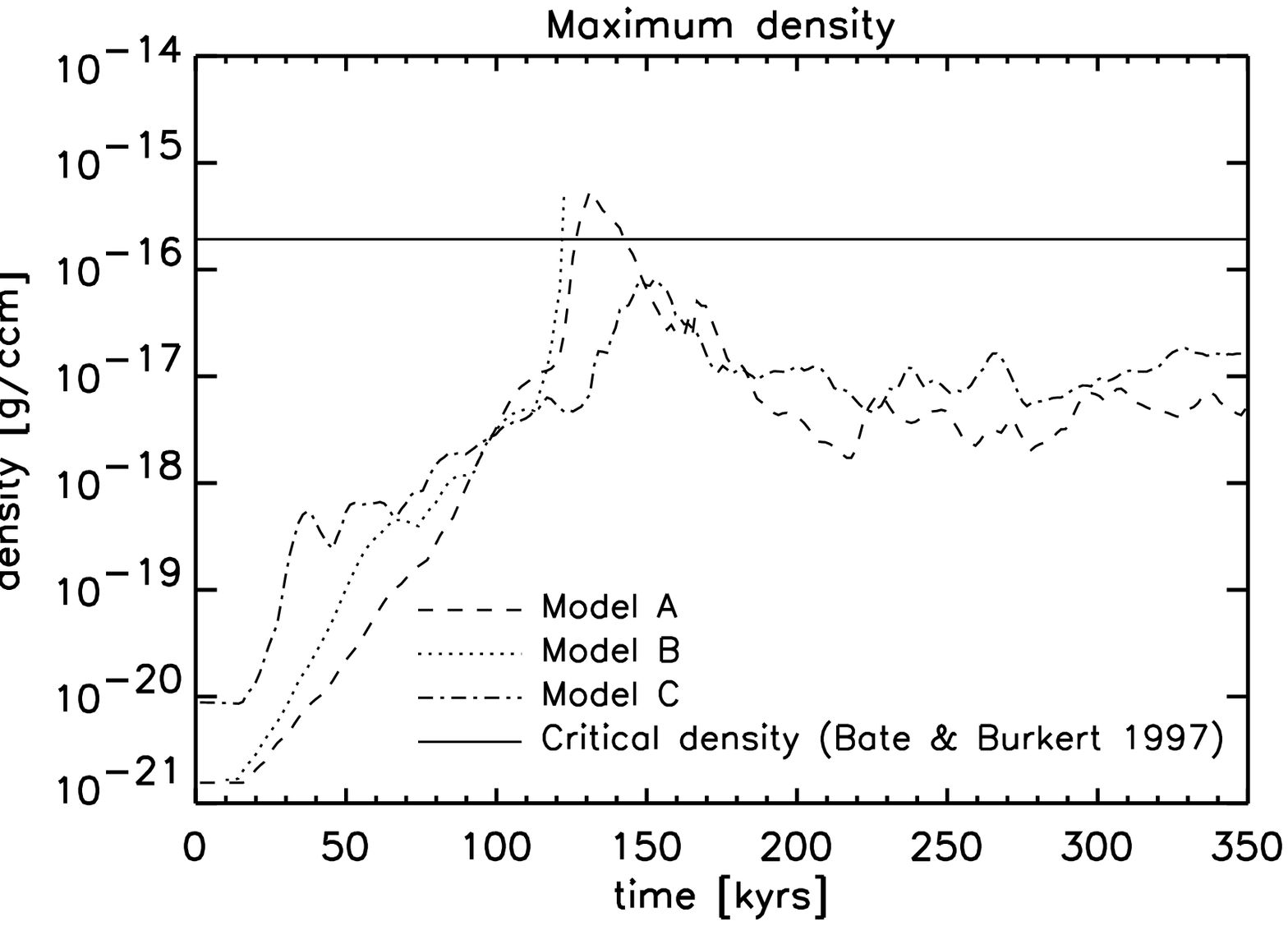,width=8.5cm}
    \epsfig{file=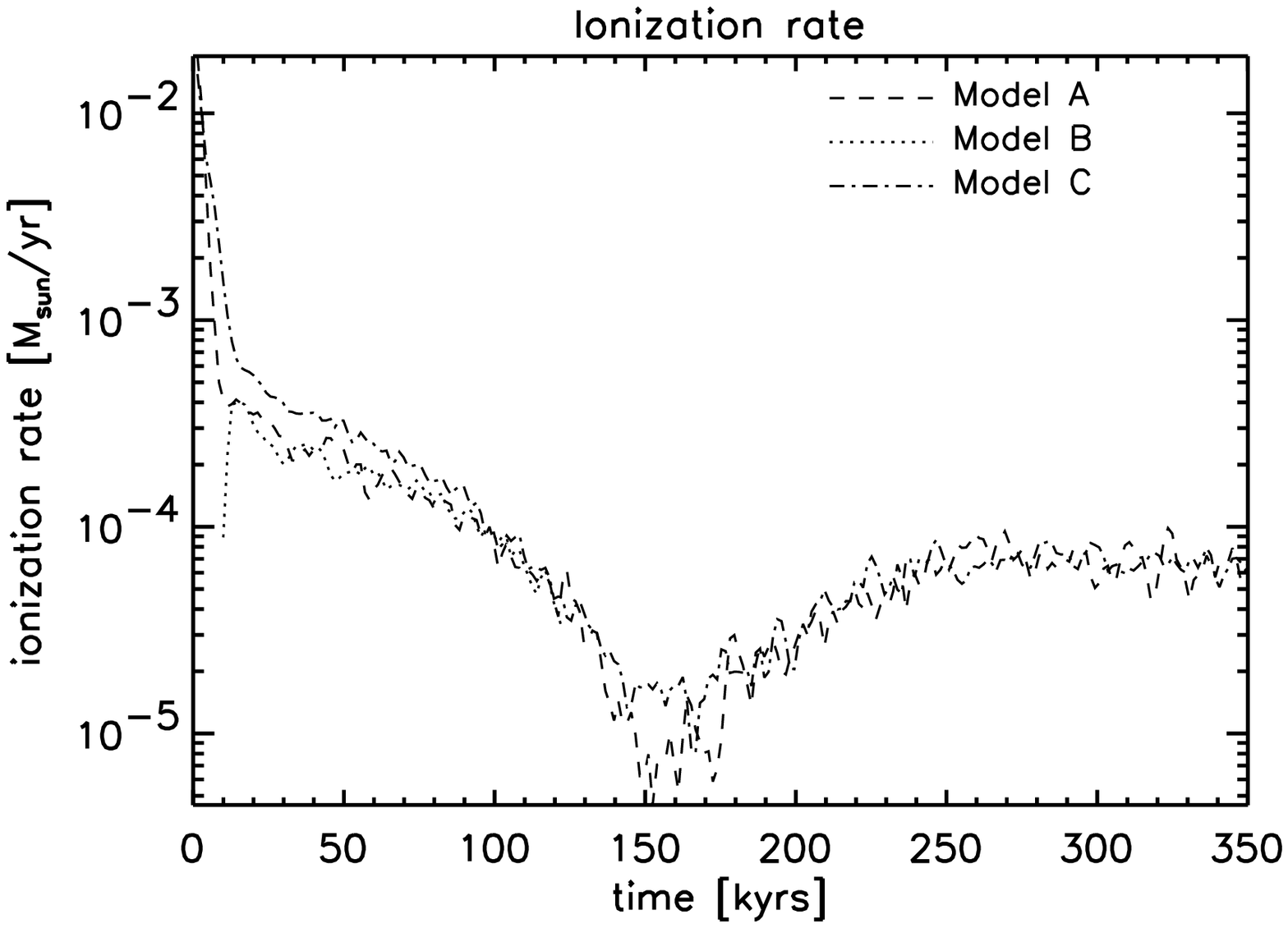,width=8.5cm}
  \caption{Maximum density (upper panel) and ionization rate (lower
    panel) in the systems. One can distinguish between
  the four different phases mentioned in Sect.~\ref{sect:caseA}. The
  dense shocked layer builds up for $t<130$~kyrs. At $t\simeq
  130$~kyrs, the systems reach maximum compression. Here model B suffers gravitational collapse, which reveals as the steep
  rise in density. For models A and C, after a short
  reexpansion phase there follows a fairly quiescent phase, in which
  material is being constantly ionized from the surface. This becomes evident in the lower panel. Note the solid line in the upper panel, showing the critical maximum density which the numerical method can handle in the cases with gravity (Bate \& Burkert 1997).} \label{fig:maxdens} 
\end{figure}

Placing our models with $R_{\rm pc}^2/S_{49} \simeq 4.71$ and $m_\odot
\simeq 40$ into the Figure~14 in Bertoldi \&
McKee~\shortcite{bertoldi90} (reprinted here as Fig.~\ref{fig:bertoldi}), they lie far in the regime where gravitational instability is expected. Simulations with smooth initial conditions including self-gravity verify this conclusion. The overall evolution is indistinguishable to the one in case~A, until the point of maximum compression is reached. At that moment gravitational collapse sets in, which can be seen in the sudden rise in the maximum density of the system (see Fig.~\ref{fig:maxdens}).

\subsection{Time evolution with self-gravity, large-scale and small-scale initial density perturbations: cases C and D}

In these models, the evolution fundamentally differs from the one in model B. During the compression phase, density enhancements are amplified in
the shocked layer, similar to what is observed in the simulations
by Elmegreen et al.~\shortcite{elmegreen95} (Figs. \ref{fig:C1}, \ref{fig:C2})
and those of Williams et al. (2001) (Fig. 6f). 
They are reminiscent of the observed small-scale structures seen in M16
(Hester et al. 1996; White et al. 1999). Contrary to case B, no collapse occurs at the
moment of maximum compression. We rather observe an evolution
analogous to the phases 3 and 4 of case A. After a delay of 300 kyrs, the layer
starts fragmenting by gravitational instability.

\section{Discussion} \label{sect:discussion}

\subsection{Jeans Criterion}

As already pointed out in Chapter~\ref{chapt:inconds}, the initial conditions were chosen such that the integration volume contains $\sim 1$ Jeans mass. Looking at the Jeans mass $M_{\rm J}$ and Jeans length $L_{\rm J}$,
\begin{equation}
  M_{\rm J} = \frac{4}{3} \pi^{5/2} \left( \frac{kT}{Gm} \right)^{3/2} \rho^{-1/2}; L_{\rm J} = \sqrt{\frac{15 kT}{2 \pi \rho G m}},
\end{equation}
and the freefall timescale,
\begin{equation}
  T_{\rm ff} = \sqrt{\frac{3 \pi}{32 G \rho}},
\end{equation}
it follows that $M_{\rm J}$ as well as $T_{\rm ff}$ drop with
increasing density. At the moment of highest compression in model A, regions with densities higher than a threshold density $\rho_{\rm m} = 5 \cdot 10^{-17} {\rm g \, cm}^{-3}$ contain $5.3\; M_\odot$. For this $\rho_{\rm m}$, one finds $M_{\rm J} = 2.7 \, {\rm
M}_\odot$, $L_{\rm J} = 7 \cdot 10^{-3}\, {\rm pc}$ and $T_{\rm ff} = 10 \, {\rm kyrs}$. Compared to these values, the compressed region in the tip of the cloud contains 2 Jeans masses and is large enough, and $T_{\rm ff}$ is small compared to other timescales in the system. One can thus expect that the globule is unstable to gravitational collapse at this moment. This explains why we observe collapse when including gravity in model B.

One should however use the Jeans masses criterion with care, as the cases C and D illustrate. It only considers the balance between internal energy and gravity and is thus
a static approach, neglecting the effects of internal motion. It is
as such only applicable in very few cases.

\subsection{Reason for different behaviour of case B with respective to cases C and D}

Since the initial conditions are chosen such that they contain
$\sim 1$ Jeans mass for all cases, the stabilization of cases C and D cannot
be caused by gas pressure alone. The only mechanism which
differs in the simulations and which can be responsible is
undirected gas motion, i.e. disordered kinetic energy.

\begin{figure}
    \epsfig{file=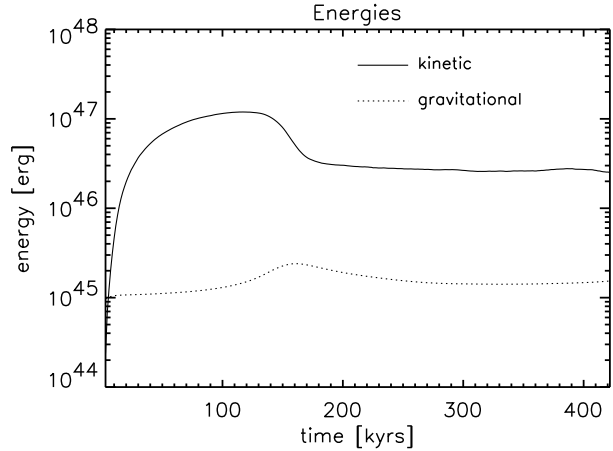,width=8.5cm}
  \caption{Energies vs. time in the neutral gas for model C. See text for details.} \label{fig:energies} 
\end{figure}

Fig.~\ref{fig:energies} shows the total amount of kinetic
energy $E_{\rm kin}$ relative to the center of mass compared to the
gravitational energy $E_{\rm grav}$ in the system for case C. During the
compression phase, more and more material is collected by the
post-shock layer and accelerated to velocities $\sim 5$~km/s. As a
consequence, $E_{\rm kin}$ grows steadily until the shock converges
along the axis of the configuration. Then follows a short phase of
strong dissipation caused by the colliding shocks. During reexpansion
and acceleration of the globule, there is still a residual amount of
$E_{\rm kin}$ left in the globule.

The difference between cases C and D with respect to model B is that in
the latter case the gas flow is strongly converging in the tip of the
globule, whereas in the former the shock front is refracted by the
density fluctuations and thus a substantial amount of $E_{\rm kin}$ is
transformed into undirected motion on scales small enough to support the globule. Additionally, the knots formed in the compression layer are
relatively stable entities, which behave like a swarm of bullets. They
tend to pass by each other without merging. Both effects inhibit
gravitational collapse during maximum compression for cases C and D. In 2D
calculations using cylindrical symmetry, these bullets would be represented
by concentric rings and their cross section for collision would be artificially enhanced. It is thus doubtful whether the delay could be observed at all in 2D calculations.

In the late phases in models C and D, all models show gravitational instability in the dense layer which was built up in the tip of the cloud facing the source, which leads to fragmentation and disruption of the cloud (Fig. \ref{fig:C2}, lowest panel). This happens at different evolutionary times for several models we calculated, varying between 420~kyrs and 600~kyrs. We don't show these late stages in Fig.~\ref{fig:maxdens}, since then over 90 per cent of the gas is already ionized. This means that numerical resolution is poor due to low particle numbers in the neutral gas. A second reason why we doubt the reliability of the time when instability occurs is that it will strongly depend on the dissipation rate of the undirected kinetic energy. Since our numerical scheme uses artificial viscosity, we expect numerical dissipation which cannot represent the real value. However, the important result we trust in is the delay in gravitational collapse after maximum compression which we observe in all models with perturbed initial conditions.
    
The delay in the collapse for models C and D was neither observed nor
anticipated in earlier work. Bertoldi \& McKee~\shortcite{bertoldi90}
derived the conditions for which gravitational collapse can be
expected by analysing the balance between the gas pressure in
stratified clouds and the external pressure and recoil. In their
Figure~14, reprinted here as Fig.~\ref{fig:bertoldi}, all the models presented here fall far into the unstable
region. Note that our globules are far from stratified; the timescale
for stratification is roughly given by the sound crossing time,
\begin{equation}
t_{\rm sc}=2r/a_{\rm c},
\end{equation}
where $r$ is the radius of the globule and $a_{\rm c}$ the sound
velocity in the neutral gas. In our case,
\begin{equation}
t_{\rm sc} \simeq \frac{0.5 \, {\rm
pc}}{0.3 \, {\rm pc}/{\rm Myrs}} \simeq 1 {\rm Myr},
\end{equation}
which is substantially longer than the timescales considered here.  

\subsection{Properties of the ``bullets'' in case C}

\begin{figure}
    \epsfig{file=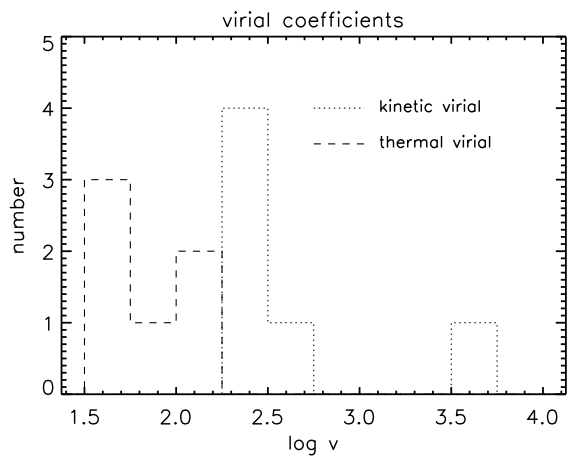,width=8.5cm}
  \caption{Kinetic and thermal virial coefficients for the six most massive bullets in model C at t=100 kyrs.} \label{fig:virial_hist}
\end{figure}

In Fig.~\ref{fig:virial_hist}, the distribution of thermal and kinetic
virial coefficients  for the six most massive ``bullets'' in case
D are shown after 100~kyrs evolutionary time, shortly before maximum compression. The coefficients were
estimated by taking the ratio of internal or kinetic energy, respectively,
compared to the gravitational energy at a radius, where the bullet is
by its higher density still well separated from the embedding
medium. The virial coefficients are all well above 10, which indicates
that the bullets are no gravitationally bound entities, but are
pressure-confined. This assumption is even affirmed due to the fact
that all of the bullets are situated in the shocked, dense layer which
is formed by the overpressure in the neighbouring ionized medium and
the recoil caused by gas traversing the ionization front with high
velocity.

The hydrodynamic lifetime of the bullets, estimated from their extent
and the sound velocity in the neutral medium, are found to lie between
100 and 400~kyrs, which is comparable to the time the shock front
needs to penetrate the whole globule. 

\subsection{Simplifications and caveats of the models}

\subsubsection{Equation of state}

In the calculations presented above, we use a very simplified equation of state. We assume that the heating and cooling timescales are short compared to the other important timescales, and that the equilibrium temperature is constant at $T_{\rm c}=10 {\rm K}$ in the neutral and $T_{\rm i} = 10\,000 {\rm K}$ in the ionized regions. For material in the ionization front, we couple the temperature to the ionization fraction by
\begin{equation}
  T = T_{\rm c} \cdot (1-x)+T_{\rm i} \cdot x.
\end{equation}
This ansatz is unsatisfactory in several respects. First, the
temperature of recombination zones is severely underestimated, since
the temperatures of recently recombined gas tend to lie in the regime
of several thousand degrees Kelvin. We expect that this leads to
overestimated densities in the shadow region ``behind'' the main
clump. We thus expect the mass reservoir to be overestimated as well, leading to longer lifetimes and increased probability for gravitational collapse in the late stages after rebounce.

It would be desirable to implement a detailed heating and cooling
model, which takes into account the temperatures of dust and gas
separately. This implies the application of a more general method for
determining the radiation field generated by the massive stars. One
possibility is application of flux-limited diffusion, which was
proposed by Levermore \& Pomraning~\shortcite{levermore} and already successfully implemented into grid codes \cite{yorkekaisig}. This issue is currently being addressed.

\subsubsection{Diffuse recombination field}

The model used for solving the radiation transfer of ionizing photons uses the simplification called ``on the spot assumption''. Lyman continuum photons produced by electrons falling into the ground state of protons are themselves capable of ionizing H atoms. We assume that this happens very close to the site of their production and that following the exact equations of radiation transfer for these photons can be neglected. It was shown by Yorke et al.~\shortcite{yorkewelz} that this assumption is not valid in special cases where the diffuse recombination radiation leads to the penetration of ionization fronts into regions which are shielded from the direct illumination by the ionizing stars. Those regions can thus be heated to temperatures around $3000\;{\rm K}$. As in the paragraph above, this again affects the regions where ionized gas streams into the shadows and starts recombining (see for example Fig.~\ref{fig:C1}, lower panels).

We expect that the recombination of this material will be delayed, if not prevented, by the Lyman continuum photons. This can also be addressed as soon as an implementation of the Flux-Limited Diffusion Approximation for SPH is at hand. 

\section{Summary}

In this paper we demonstrate that the physics of Radiation Driven Implosion strongly depends on the properties of the initial cloud material. In the presence of initial perturbations, the evolution is highly nonlinear and the fate of the imploding cloud is not easy to foresee. We find that gravitational collapse need not necessarily occur in those objects, depending on the presence, size and strength of initial density perturbations. The redistribution of kinetic energy from the implosion into undirected motion can stabilize the globules for a significant amount of time. Unfortunately, it is difficult to estimate this delay from the models presented here, since it depends on the decay rate of the kinetic energy, and the SPH method used here can reproduce this rate only insufficiently.

The delay could be relevant for star formation if the globule is
stabilized for such a long time that the ionizing radiation can
totally erode the globule. If gravitational collapse occurs before, we
expect the collapsing regions to decouple from the gas around
them. The remaining gas will then be further accelerated by the rocket
effect from the ionization front, whereas the decoupled cores will lag
behind and eventually be overtaken by the ionization front. This
process could then repeat for several times, as long as there is still
mass in the reservoir. We expect the outcome to be very similar to the
observations of Sugitani et al.~\shortcite{sugitani99,sugitani00},
where young stellar objects are aligned to the axis of the globules
with an age gradient from the oldest objects far away from the
globules and the youngest ones still sitting inside. By using the
numerical technique called ``sink particles''~\cite{bate}, we are
confident to be able to reproduce this scenario. One of the next steps
should be to take into account the measured internal motion in
molecular clouds. We will present calculations with clouds initially supported by turbulent motion in a subsequent paper.

\section*{Acknowledgments}
This work was supported by the Deutsche Forschungsgemeinschaft (DFG),
grants Bu 842/4-1 and Bu 842/4-2. The 3D projections in Figs. \ref{fig:A1} -- \ref{fig:C2} were computed with the help of {\it starsplatter} by Joel Welling and John Dubinski (http://www.psc.edu/Packages/StarSplatter\_Home/).

\end{document}